\documentclass[twocolumn]{aastex631}
\usepackage{mathptmx}

\shorttitle{Aligned binaries}
\shortauthors{Marcussen et al.}
\begin{document}

\title{THE BANANA PROJECT. VI. Close double stars are well aligned with
noticeable exceptions; results from an ensemble study using apsidal
motion and Rossiter-McLaughlin measurements.}

\author[0000-0003-2173-0689]{Marcus L.\ Marcussen}
\affil{Stellar Astrophysics Centre, Department of Physics and Astronomy, Aarhus University, Ny Munkegade 120, 8000 Aarhus C, Denmark}
\email{marcuslmarcussen@gmail.com}

\author[0000-0003-1762-8235]{Simon H.\ Albrecht}
\affil{Stellar Astrophysics Centre, Department of Physics and Astronomy, Aarhus University, Ny Munkegade 120, 8000 Aarhus C, Denmark}

\begin{abstract}
Here we present an ensemble study of spin-orbit alignment in 51 close double star systems. We determine spin-orbit angles, obliquities, in 39 of these systems making use of recently improved apsidal motion rate measurements and apsidal motion constants. In the remaining 12 systems researchers have constrained spin-orbit alignment by different combinations of measurements of apsidal motion rates, projected obliquities and stellar inclinations. Of the 51 systems 48 are consistent with alignment albeit with some measurements having large uncertainties. A Fisher distribution with mean zero and a concentration factor $\kappa = 6.7$ represents this ensemble well. Indeed employing a bootstrapping resampling technique we find our data on these 48 systems is consistent with perfect alignment. We also confirm significant misalignment in two systems which travel on eccentric orbits and where misalignments have been reported on before, namely DI Her and AS Cam. The third misaligned system CV Vel orbits on a circular orbit. So while there are some glaring exceptions, the majority of close double star systems for which data are available appear to be well aligned.

\end{abstract}

\section{Introduction} \label{sec:intro}

Stars in double star systems might be expected to have their rotation axes parallel to their orbital axes, a trait that is thought of as a natural consequence of star formation. This is because the direction of the orbit and rotation of stars is determined by the angular momentum of the molecular cloud from which they form. However a number of mechanisms have been proposed which might lead to misalignment, even in close systems. A chaotic formation history \citep{bate2010,Thies+2011,fielding2015}, or torques from a warped circumbinary disk \citep{AndersonLai2021} have been suggested. Alternatively, processes happening after stellar formation may also cause spin-orbit misalignment, e.g.\ the Kozai-Lidov mechanism which can occur if an inclined third body orbits further out \citep[e.g.][]{fabrycky2007,naoz2014,AndersonLaiStorch2017}. 

However, observational data remain sparse. Spin-orbit alignment has been studied in wide double star systems with several au separations \citep[e.g.][]{weis1974,hale1994,glebocki1997,howe2009,JustesenAlbrecht2020}, but \cite{JustesenAlbrecht2019} showed that with the data at hand no general statement about alignment in such wide systems can be made. A bit more is known about closely orbiting systems with separations less than one au. Specifically a few misalignments have been reported, namely DI Her \citep{albrecht2009}, AS Cam \citep{pavlovski2011}, and CV Vel \citep{Albrecht2014CVVel}. AI Phe may be another misaligned system \citep{sybilski2018} although these authors caution that their measurement might require additional observations to confirm this misalignment. For that reason we do not include the system in this study.

Determining the distribution of stellar obliquities among binaries and any trends herein may improve our understanding of double stars, their formation and evolution. In section \ref{sec:method} we describe how we use measurements of the rate of change of the argument of the periapsis $\dot \omega$ (or simply apsidal motion or apsidal precession rate) of binaries to constrain their spin-orbit misalignment \citep{Sterne1939,Mazeh2008}. Here we also describe the obliquity data we have on additional systems. In section \ref{sec:single systems} we present the resulting posteriors for all 51 systems of the sample and discuss individual systems we find to be inconsistent with alignment. We then perform an ensemble study of alignment as developed for exoplanets by \cite{fabrycky_winn2009} in section \ref{sec:ensemble} and summarize our findings in section \ref{sec:discussion}.

\section{Our method}  \label{sec:method}

A number of methods exist with which obliquities in double star systems can be accessed. In eclipsing systems observations of the Rositter-McLaughlin (RM) effect \citep[e.g.][]{Struve1931,DryomovaDryomov+2019}, gravity darkening \citep[e.g.][]{barnes2009,Ahlers+2020}, and spot crossing \cite[e.g.][]{Desert+2011,dai_et_al2018} might be used. For non eclipsing systems, interferometry \citep[e.g.][]{petrov1989,lebouquin2009}, asteroseismology \citep[e.g.][]{GoughKosovichev1993,chaplin_miglio2013},  or the so called $v \sin i$ method \citep[e.g.][]{schlaufman2010,JustesenAlbrecht2019} can be employed. Each of these different methods are applicable to subsets of systems with specific parameters, e.g., distance, brightness, stellar, and orbital parameters, and some require large amounts of observing time and/or space based data. Nevertheless, information on alignment has been obtained for a number of binary systems, mainly using the RM effect as well as gravity darkening. 

Here we complement these direct measurements using a different method. To constrain the obliquity in binaries we use observations of their apsidal motion rates along with precise determinations of their stellar parameters. The premise of the method is that the equatorial parts of rotationally flattened stars with parallel spin and orbital axes are closer - and thus the gravitational forces stronger - than that of two orbiting perfect spheres. This results in positive corrective terms to the gravitational force that depend on higher orders of their mutual distance than the usual $r^{-2}$ in turn leading to a prograde apsidal motion of the eccentric orbit \citep{Sterne1939}. In the other extreme where the rotation axes of the stars are located in the orbital plane of the binary, rotational flattening causes mass to be "missing" from the orbital plane and the corrective terms are negative, leading to a retrograde contribution to the apsidal motion. This method is therefore applicable to binaries whose components are rotationally flattened enough to cause a significant change to the apsidal motion rate. 

Measurements of apsdial motion rates have in general agreed well with theoretical calculations and allowed for the study of the radial mass distribution inside stars \citep{ClaretGimenez1993}. However, DI Her and AS Cam are two binary systems where the apsidal motion rates have long been known to be lower than the values predicted by theory - assuming spin-orbit alignment \citep{GuinanMaloney1985,Khaliullin1983}. A number of different explanations had been proposed to account for the discrepancy, among these a failure of General Relatively. See \cite{claret1998} for a summary of proposed explanations for the DI Her system.  
\cite{albrecht2009} established that both components of DI Her are substantially misaligned to their orbit thereby finding that the proposal by \cite{shakura1985}  \citep[see also][]{GuinanMaloney1985,CompanyPortillaGimenez1988,ReisenbergerGuinan1989} was the correct solution: misaligned stellar spin axes to the orbital axis caused the lower than expected apsidal motion rate. The cause of the misalignment in DI Her is not yet known. 
A low apsidal motion rate for the AS Cam system was first observed by \cite{Khaliullin1983}. Recently \cite{kozyreva2018evolution} found some indirect evidence pointing towards the presence of a third-body. Such a potential third body could change the orbital elements of the close binary including the position of the periastron. Another hypothesis was investigated by \cite{pavlovski2011}, who point out that the discrepancy between theoretical values and observations can be resolved if both components of the system have a significant spin-orbit misalignment also along the line-of-sight. 

In this work we take a similar approach to the one taken by \cite{pavlovski2011}. See also \cite{PhilippovRafikov2013} for a similar study on DI Her. We attempt to reverse the approach that has traditionally been taken: 

Traditionally a priori spin-orbit alignment is assumed and GR as well as the radial mass distribution inside stars, as obtained from stellar models, is tested via the internal structure constant. Which physics can be tested does depend on the parameters of the system in question (see section~\ref{sec:apsidal}). Here we assume GR to be true and that the internal structure constants for the stars in our ensemble are indeed known to the extent their formal uncertainties suggest. We are not aware of detections of third bodies in any of the systems and assume here that $\dot \omega$ is not influenced by such a body.\footnote{For the systems CW Cep, V539 Ara and Zeta Phe which do have companions \cite{Claret+2021} did calculate the apsidal motion rate caused by these bodies and found them to be small compared to the total apsidal motion rate found in these systems.} % at least to the extent that these would influence $\dot \omega$ more than its measurement uncertainty. 
With these assumptions we then attribute any discrepancy between the observed and expected apsidal motion rate to a misalignment of stellar spin axes, or if agreement is found, determine upper limits on any potential obliquity.  The recent high-precision apsidal motion determinations by \cite{baroch2021} and \cite{Claret+2021} using TESS light curves sparked this idea and their data compose the bulk the apsidal motion rates that we use. 

\subsection{Geometry}
To describe the geometry of the problem we follow the nomenclature established by \cite{fabrycky_winn2009}. The observer-oriented coordinate system we use is defined as follows: $\hat Z$ points towards the observer and $\hat X$ and $\hat Y$ lie in the sky-plane. The coordinate system is oriented such that the axis of orbital angular momentum (henceforth simply orbital axis) lies in the $\hat Y\hat Z$ plane. The orbital inclination $i_o$ is then the angle between the $\hat Z$-axis and the orbital axis. The stellar inclinations of the stars of a binary system, that is the angle between $\hat Z$ and their rotation axes, we denote $i_{s,j}$ where $j = 1,2$ refers to the primary and secondary components respectively. The stellar obliquities, which are the angles of misalignment between the orbital and stellar rotation axes, we denote $\psi_1$ and $\psi_2$.

\subsection{Obliquity from apsidal motion}
\label{sec:apsidal}

\begin{figure*}
\begin{center}
\includegraphics[trim= 4.5cm 4.3cm 4.5cm 4.5cm, clip,width=1.0\textwidth]{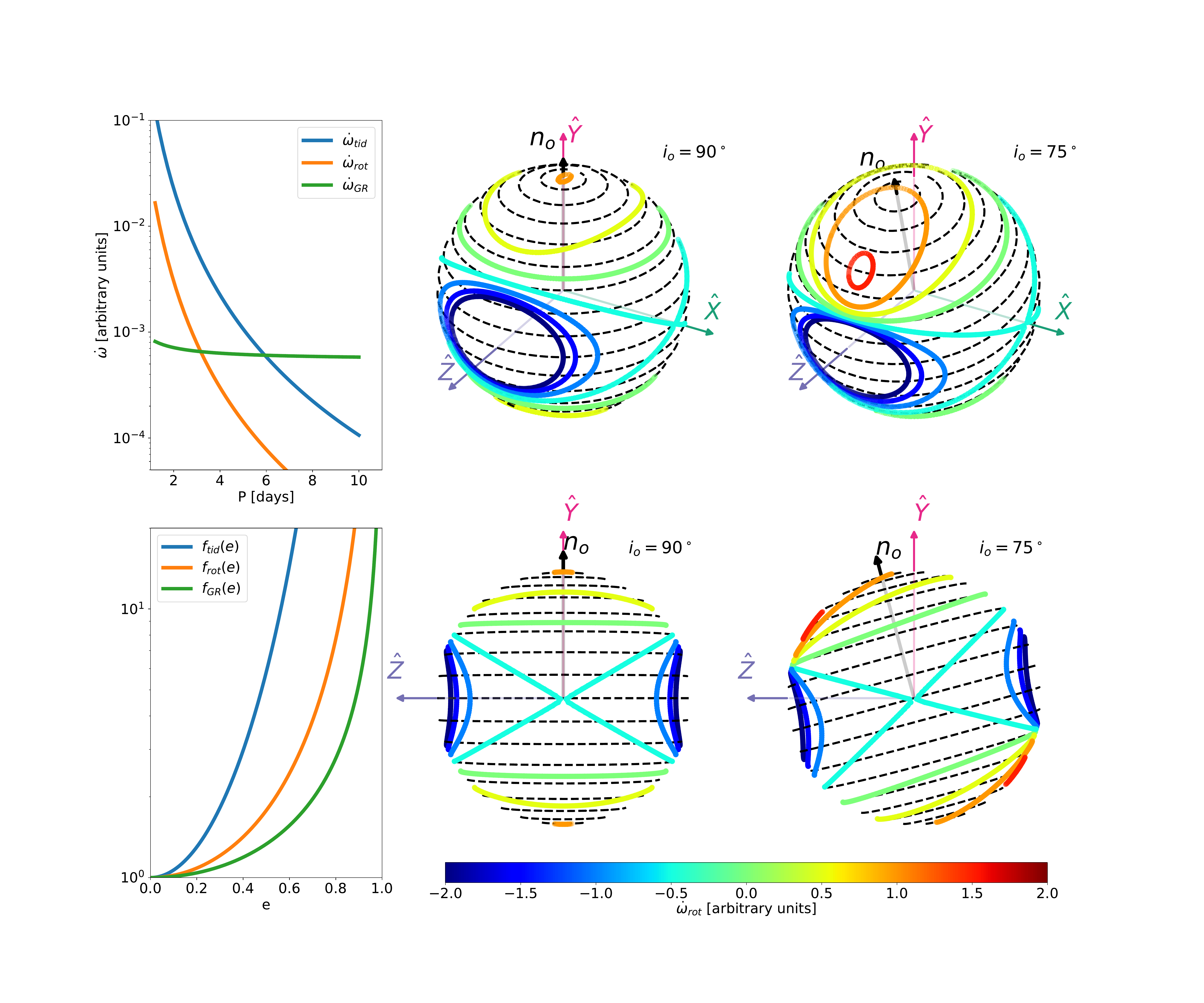}
\end{center}

\caption{{\bf Apsidal motion rate dependencies.} {\it Left top panel}: The relative strengths of the General Relativity, tidal and rotational terms of equation \ref{eq:omegaTotal} are compared in the top left panel using $e = 0$, $M_1 = M_2$ and $\Omega_j = \Omega_o$. {\it Left bottom panel}: The dependency on the eccentricity of these three terms (equations \ref{eq:GRe}, \ref{eq:tide} and \ref{eq:rote}) are compared. {\it Top and bottom middle panels}: The relative size of $\dot \omega_{rot}$ as a function of the orientation of the stellar spin axis in the observer oriented coordinate system. The contour lines of $\dot \omega_{rot}$ correspond to the values $-2,-1.5,-1,-0.5,0,0.5,1,1.5$ and $2$ and are scaled such that $\dot \omega_{rot} = 1$ when $\psi = 0^\circ$ and $i_o = 90^\circ$. The black dashed contour lines, originating from the orbital axis $\mathbf{n_o}$, represent increments of $10^\circ$ of the obliquity $\psi$. {\it Right hand panels}: The same as in the middle panels but for $i_o = 75^\circ$. The dependency of $\dot \omega_{rot}$ on both $\psi$ and $i_s$ (because $\Omega_j$ is inferred from an observed $v\sin i_s$) is revealed because the contour lines of $\dot \omega_{rot}$ do not follow the contours of $\psi$. \label{fig:dependencies}}
\end{figure*}

The apsidal motion rate of a binary system $\dot{\omega}$ is the sum of contributions from General Relativity, $\dot{\omega}_{GR}$, classical gravitational interactions caused by the asphericity of the constituent stars, $\dot{\omega}_{cl}$ as well as potential contributions from third bodies \citep{shakura1985}. Unless confirmed in other studies, we set this term to zero. An asphericity of a star may arise from the deformations caused by either tidal interactions between the components of a binary system or from rotational flattening that leads to a larger diameter of a star along its equatorial plane than its north-south axis. The total apsidal motion rate may thus be expressed as \citep{shakura1985,baroch2021}:

\begin{equation} \label{eq:omegaTotal}
    \dot{\omega}  = \dot{\omega}_{GR} + \dot{\omega}_{tid} + \dot{\omega}_{rot},
\end{equation}

where each term, in units of degrees per cycle, are:

\begin{equation} \label{eq:omegaGR}
    \dot{\omega}_{GR}  = 5.447 \times 10^{-4} \frac{(M_1 + M_2)^{2/3}}{P^{2/3}} f_{GR} (e)\\
\end{equation}

\begin{equation} \label{eq:omegatid}
    \dot{\omega}_{tid}  = 360 \times \sum^2_{j=1} 15 k_{2,j}  r^5_j \frac{M_{3-j}}{M_j} f_{tid}(e)
\end{equation}

\begin{equation} \label{eq:omegarot}
    \dot{\omega}_{rot} = 360 \times \sum^2_{j=1} k_{2,j} r^5_j \left(1 + \frac{M_{3-j}}{M_j} \right)\left( \frac{\Omega_j}{\Omega_o} \right)^2 f_{rot}(e)\phi_j (\psi_j)
\end{equation}

where $P$ is the orbital period, $M_j$ are the stellar masses, $r_j = R_j/a$ are ratios between radii $R_j$ and the orbital semi-majoraxis $a$. $\Omega_j = v_j/R_j$ are angular rotation velocities, where $v_j$ is the rotation velocity of component $j$ and $\Omega_o = 2\pi/P$ is the angular orbital velocity. Finally, $k_{2,j}$ are the internal structure constants (or apsidal motion constants) that relate to the radial mass-distribution of the stars \citep{Kopal1978} and are determined from measurements of the apsidal motion \citep{ClaretGimenez1993}. The three terms $\dot{\omega}_{GR}$, $\dot{\omega}_{tid}$ and $\dot{\omega}_{rot}$ are all increasing functions of the eccentricity $e$ according to their respective functions $f_{GR}$, $f_{tid}$ and $f_{rot}$ \citep{Sterne1939,baroch2021}, see also fig.~\ref{fig:dependencies}:

\begin{equation} \label{eq:GRe}
    f_{GR}(e) = \frac{1}{1-e^2}
\end{equation}

\begin{equation}\label{eq:tide}
    f_{tid}(e) = \frac{1 + 1.5e^2 + 0.125e^4}{(1-e^2)^5}
\end{equation}

\begin{equation}\label{eq:rote}
    f_{rot}(e) = \frac{1}{(1-e^2)^2}
\end{equation}

Whereas $\dot{\omega}_{GR}$ and $\dot{\omega}_{tid}$ both solely have positive (prograde) contributions to the apsidal motion rate, the sign and size of $\dot{\omega}_{rot}$ depends on the stellar obliquity. This dependency is expressed by the function $\phi_j (\psi_j)$, which as per \cite{shakura1985}, is given by

\begin{equation} \label{eq:phi}
    \phi_j (\psi_j) =  \frac{ 5\cos^2\psi_j - 1}{2} -\frac{\cos{\psi_j}\left(\cos{\psi_j} - \cos{i_{s,j}\cos{i_o}} \right)}{\sin^2 i_o}
\end{equation}

It may seem surprising that $\phi$ should depend on the orientation of the observer (via the terms $i_o$ and $i_{s,j}$) but this comes about because of the precession of the orbital axis around the stellar axes that occurs in addition to apsidal precession when $\psi \neq 0^\circ,90^\circ$, making the longitude of the ascending node $\Omega$ change over time. This in turn shifts the zero-point from which $\omega$ is measured, effectively translating into a change over time of $\omega$ if also $i_o \neq 90^\circ$.\footnote{We provide a visualisation program which encompasses the apsidal motion and axial precession of the orbital axis around the stellar spin axis given all the contributions from \ref{eq:omegaTotal} under this url: \href{https://phys.au.dk/exoplanets}{https://phys.au.dk/exoplanets}. The system parameters including the stellar spin rate and orientation can be manipulated. The program runs under Linux, Windows, MacOS as well as in a VR environment. Comments and suggestions might be provided to the lead author.}

$\psi_j$ may in turn be expressed as a function of $i_o$, $i_{s,j}$ as well as the projection of $\psi_j$ unto the sky-plane, $\lambda_j$ \citep{fabrycky_winn2009}:

\begin{equation} \label{eq:psi}
    \cos \psi_j = \sin i_{s,j} \sin i_o \cos \lambda_j + \cos i_{s,j}\cos i_o.
\end{equation}

In the left panel of fig.~\ref{fig:dependencies}, the relative strengths of the GR, tidal, and rotational terms are compared, using $M_1 = M_2$ and $\Omega_j = \Omega_o$. To the right of these, the dependence of $\dot \omega_{rot}$ on the orientation of the stellar spin axis in the observer-oriented coordinate system is shown. Although not explicitly shown in equation \ref{eq:rote}, for systems where only the sky-projection of the rotation speed of a star $v \sin i_s$ and not its true rotation speed $v$ is known, $\dot \omega_{rot}$ rises rapidly towards lower $i_s$ because $v$ and by extension $\Omega_j$ is determined through

\begin{equation}
    v = \frac{v \sin i_s}{\sin i_s}.
\end{equation}

When determining $\psi_j$ from a system's apsidal motion alone, there is an intrinsic degeneracy between $\psi_1$ and $\psi_2$. For all but the few systems where $\lambda_j$ are known, the two stellar rotation axes are therefore taken to be parallel and the obliquity is then denoted $\langle \psi \rangle$. When the primary and secondary are similar in size, mass, and rotation speeds it follows that $\dot \omega_{rot,1} \approx \dot \omega_{rot,2}$. $\langle \psi \rangle$ can then be thought of as the mean misalignment of a system. When say the primary dominates the apsidal motion rate budget, $\langle \psi \rangle \approx \psi_1$. Another degeneracy exists between $\psi_j$ and $180^\circ - \psi_j$ because the corrections to the gravitational potential from a rotationally flattened star are the same whether the rotation is prograde or retrograde. We always assume the lower values of these two for $\psi$. 

Our ability to constrain obliquities from observed apsidal motion rates alone is usually not great even for relatively well determined systems and precise rates, primarily because of the fifth power dependence on stellar radii of $\dot \omega_{rot}$. The $1\sigma$ uncertainty of $\psi$ for a typical system of our sample is around $15^\circ-20^\circ$.

\subsection{Systems with projected obliquities}

The projection of the obliquity $\psi$ onto the sky-plane $\lambda$ has been measured from observations of the Rositter-McLaughlin effect for the following three systems, where a precise apsidal motion rate is also known: DI\,Her, EP\,Cru, and V1143\,Cyg. For these three systems \cite{Claret+2021} did test if the measured projected obliquities are consistent with the measured apsidal motion rates and find this to be the case. In addition to these systems we found another nine systems with projected obliquity measurements where the apsidal motion has not been determined. We incorporate these systems in our ensemble study as well in order to get a more complete picture of the alignment distribution of binaries in general. Stellar obliquities are already quoted for two of these nine systems: Kepler-16 and KOI-368. Using the approach of 
\cite{MasudaWinn2020} however, we re-obtain distributions of these obliquities, where we properly take into account the correlation between rotational velocities and projected rotational velocities.  

\begin{figure}
  \begin{center}
    \includegraphics[trim= 1.3cm 0.5cm 2cm 2.15cm, clip,width=1\columnwidth]{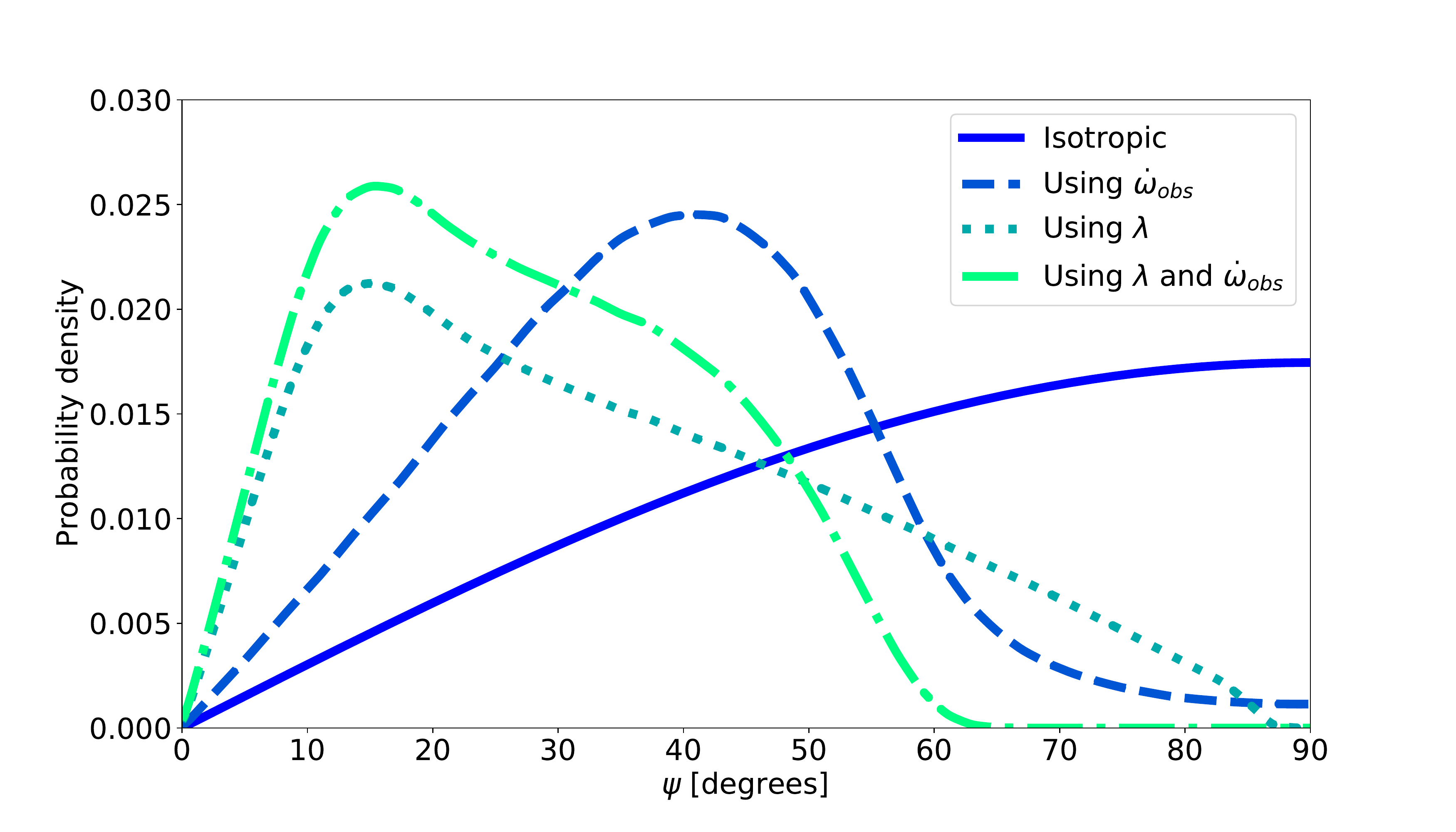}
    \caption{
    \label{fig:V1143 Cyg} 
    \textbf{Using projected obliquities}.
    A comparison of our ability to constrain $\psi$ given measurements of $\dot \omega_{obs}$ alone (dashed), $\lambda$ alone (dotted) and the combination of the two (dash dotted) as well as our prior (solid). We display the information available on V1143\,Cyg.
    }
  \end{center}
\end{figure}

In fig.~\ref{fig:V1143 Cyg} we show for the case of the V1143 Cyg system how well $\lambda$, $\dot \omega_{obs}$ and the combination of the two constrain $\psi$ compared to the uninformative, isotropic case.
By itself a measurement of say $\lambda = 5^\circ$ still allows $\psi$ to take any value between $5^\circ$ and $90^\circ$ (when $i_o = 90^\circ$) because a system may be misaligned due to a low stellar inclination. Even so, the particular probability distribution of $\psi$ is still greatly affected by such a low $\lambda$ measurement \citep{fabrycky_winn2009}. That is because a pole on configuration would then be required for $\psi$ to be large and such a pole on configuration is a priori rather unlikely, assuming an isotropic distribution (as we do). This is what leads to the drop off in probability of the dotted line in fig.~\ref{fig:V1143 Cyg}. Given that the method of constraining $\psi$ from $\dot \omega_{obs}$ is most sensitive when the stellar inclination is low (due to the high sensitivity of the theoretical $\dot \omega$ on $v$) it is  complemented well by $\lambda$ measurements.

\section{Obliquities of single systems}\label{sec:single systems}
\begin{figure*}
\begin{center}
\includegraphics[trim= 6.2cm 5.2cm 6.1cm 5.5cm, clip,width=1\textwidth]{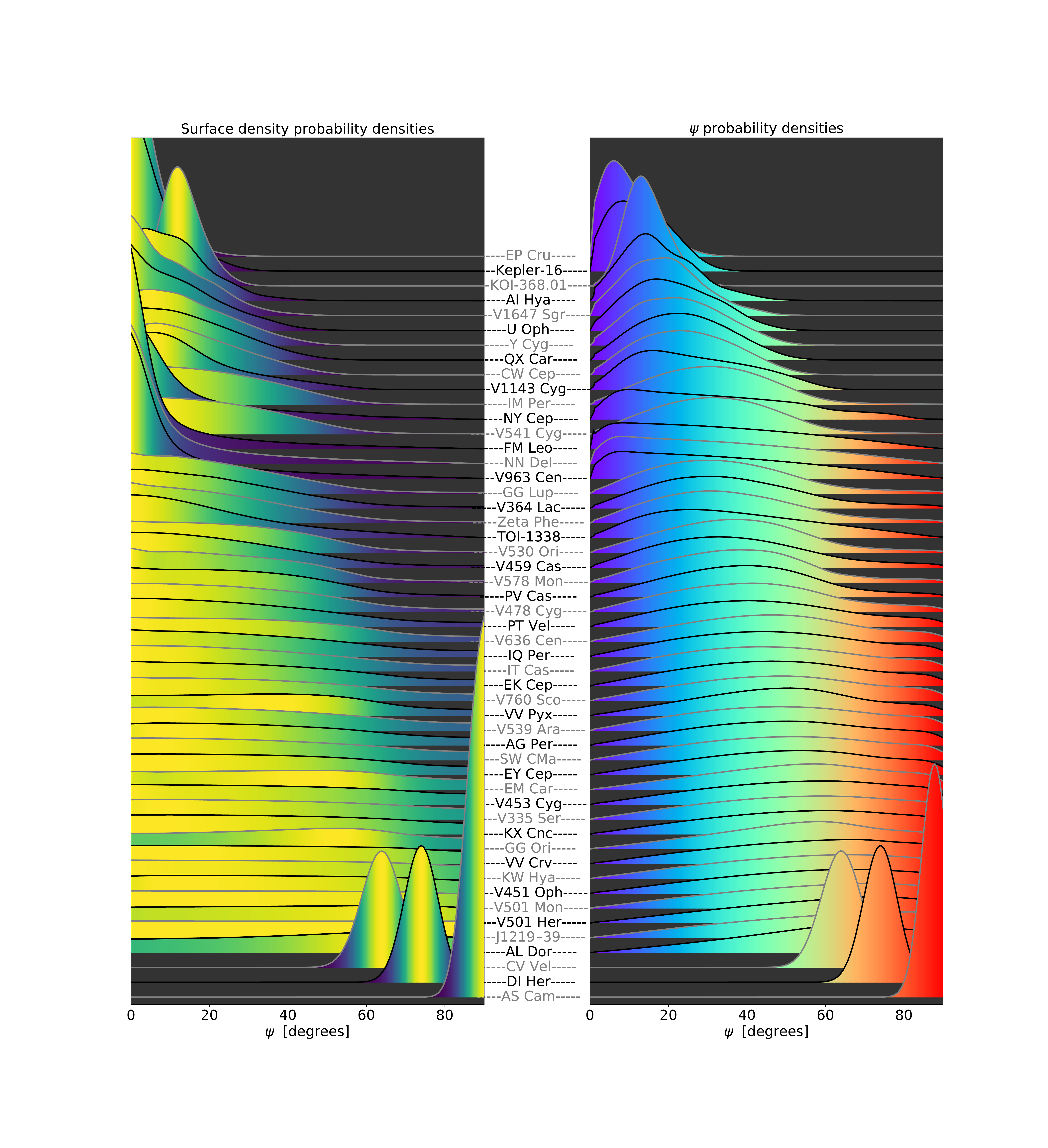}
\end{center}

\caption{{\bf Posteriors of all systems.} All systems of the sample are shown here using the same color scheme as fig.~\ref{fig:psi} for the $\psi$ probability density $p(\psi|D_n)$ and the average surface density $p(D_n|\psi)$ of a system. Although $\psi$ by itself may seem inconsistent with alignment due to a large dispersion, the surface density of the orientation of the stellar spin axis is highest where it is aligned with the orbital axis for a majority of systems as seen in the left column. \label{fig:All posterios} }
\end{figure*}

The resulting posterior probability density functions and average surface densities of $\psi$ for all 51 systems of the sample are shown in fig.~\ref{fig:All posterios}. All used system parameters and their references are listed in table \ref{tab:systems}\footnote{Table 1 by \cite{Albrecht2011NYCep} lists some earlier, qualitative results coming from RM measurements in eclipsing double stars.} along with the mode and $1\sigma$ highest posterior density intervals of the inferred $\psi$ posteriors. For some systems the apsidal motion does not provide any meaningful information about the alignment of the individual system, resulting in a close-to flat surface density probability distributions and our prior for the obliquity is recovered. For these we do not list $\langle \psi \rangle$ values. Specifically we impose the requirement that the highest density interval of the surface density must be shorter than $75\%$ of the interval a flat distribution would give, i.e.~$< 46.4^\circ$.

\subsection{Systems inconsistent with alignment}
\label{sec:oblique}

Three systems in our sample had spin-orbit misalignment reported in the literature before. These are not included in our ensemble study that we discuss in the next section (\ref{sec:ensemble}). Additionally, shortly before submission we became aware of the apsidal motion constant measurement in the triple system KOI-126 which was found by \cite{Yenawine+2021} to be inconsistent at a two sigma level with alignment. We summarize the information on the misalignment and our results for these four systems here: 

\textbf{CV Vel:} The stars in CV Vel travel on circular orbits and therefore no apsidal motion is present. We can therefore only constrain $\psi$ from its two $\lambda$ measurements, $\lambda_1 = 52\pm6^\circ$ and $\lambda_2 = -3 \pm 7^\circ$ yielding the results $\psi_1 = 57.0^{+12.5^\circ}_{8.0^\circ}$ and $\psi_2 = 13.0^{+34.0^\circ}_{-7.7^\circ}$. \cite{Albrecht2014CVVel} used additional data to constrain $\psi$ in this system however, namely changes in the projected rotation velocities over the course of 30 years and one additional $\lambda$ measurement for one component separated by several years. These changes are caused by the stellar spin precession around the total angular momentum in the system. The obliquities derived from these measurements are $\psi_1 = 64 \pm 4^\circ$ and $\psi_2 = 46 \pm 9^\circ$ which we list in table~\ref{tab:systems}. 

\textbf{DI Her:} Both $\dot \omega_{obs}$ and $\lambda$ measurements exist for DI Her. Additionally, Liang et al.\ (submitted) recently determined the obliquities of both components of DI Her using information obtained from $\lambda$, $\dot \omega_{obs}$, precession of the spin axes, as well as gravity darkening (secondary eclipse) and spot coverage (primary eclipse). They obtain $\psi_1 = 74^{+2^\circ}_{-3^\circ}$ and $\psi_2 = 80^{+2^\circ}_{-3^\circ}$. Using our approach on this system gives consistent results: $\psi_1 = 75.0 ^{+2.2^\circ}_{-4.0^\circ}$ and $\psi_2 = 81.0 ^{+4^\circ}_{-3.3^\circ}$ but we list their more precise results.

\textbf{AS Cam:} The system that resulted in the most precise single system obliquity determination is AS Cam, where a discrepancy between observed and theoretical apsidal motion rates (assuming alignment) has long been known to exist. Assuming, as we do in this work, this discrepancy to be from misalignment, we find an obliquity $\langle \psi \rangle = 86.9^{+2.2^\circ}_{-0.5^\circ}$ constrained by the apsidal motion alone. We thereby confirm the earlier result by \citep{pavlovski2011}. The maximum a posteriori solution has the stellar inclinations at $i_{s,1} = 6.5^\circ$ and $i_{s,2} = 2.1^\circ$ translating into the rotation speeds $v_1 = 128\,\mathrm{km\,s}^{-1}$ and $v_2 = 125.5\,\mathrm{km\,s}^{-1}$. This is  more in line with the expected rotation speeds of stars with effective temperatures of $12\,840 \pm 120$\,K and $10\,580 \pm 240$\,K than the projected rotation speeds of $14.5\pm0.1$~km\,s$^{-1}$ and $4.6\pm0.1$~km\,s$^{-1}$ \citep{pavlovski2011}. We see these results as supporting evidence of the misalignment hypothesis of this system.

\textbf{KOI-126:} 
While not part of our sample we would like to highlight this triple star system consisting of two M-dwarfs in a 1.76 day orbit and whose center of mass orbits an F-type star every 36 days \cite{Carter+2011}. This systems was studied in detail by \citep{Yenawine+2021} using a large body of data. These authors found a 2$\sigma$ discrepancy between the expected and observed apsidal motion constant for the inner pair of M-dwarfs. They explored the possibility that misalignment between the stellar spin axes of the two M-dwarfs and the orbital axis might have caused this disagreement. Since they have no measurements of projected rotation speeds for the M dwarfs, in order to calculate the decrease in theoretical apsidal motion rate due to a potential spin-orbit misalignment, they assume pseudo-synchronization \citep{hut1981} between $\Omega_j$ and $\Omega_o$. Given this assumption, they find that even a misalignment of $\psi_1 = \psi_2 = 90^\circ$ is not enough to account for the anomalously low $k_2$. However, we want to point out that if the assumption of pseudo-synchronization is not made, an angular rotational frequency $\Omega_j = 3.1\Omega_o$, translating into rotation speeds of $v_1 = 23$ km\,s$^{-1}$ and $v_2 = 21$ km\,s$^{-1}$, would make spin-orbit misalignments of $90^\circ$ by both components able to account for the observed discrepancy.
\section{Ensemble constraints}
\label{sec:ensemble}

Given that the obliquity of a single system (near alignment) is difficult to determine, whether by the apsidal motion method, RM effect or even the combination of the RM effect with a stellar inclination from $v \sin i_s$ and $v$, it is helpful to combine the results from many systems, which we aim to do in this section.

In the ensemble of systems for which we wish to constrain the degree of alignment we do not include the three clearly misaligned systems DI Her, AS Cam, and CV Vel. The ensemble therefore consists of 48 systems. Of these, 38 systems are constrained by their apsidal motion rate alone. 2 systems by the combination of apsidal motion and projected obliquities, 6 systems based on projected obliquities alone and finally we determined the obliquity of 2 systems from their projected obliquities and inclinations using the approach of \cite{AlbrechtMarcussenWinn+2021}.  

For this study we use a hierarchical Bayesian inference framework à la \cite{fabrycky_winn2009,munoz2018}, so a much stronger upper limit can be put on the dispersion of the ensemble distribution as a whole. In this approach it is assumed $\psi$ of all systems is drawn from a Von-Mises Fisher distribution (henceforth simply "Fisher distribution") with mean $\psi = 0^\circ$:

\begin{equation}
    p (\psi | \kappa) = \frac{\kappa}{2 \sinh{\kappa}} \exp(\kappa \cos{\psi})\sin \psi.
\end{equation}

\begin{figure*}
\begin{center}
\includegraphics[width=1.0\textwidth]{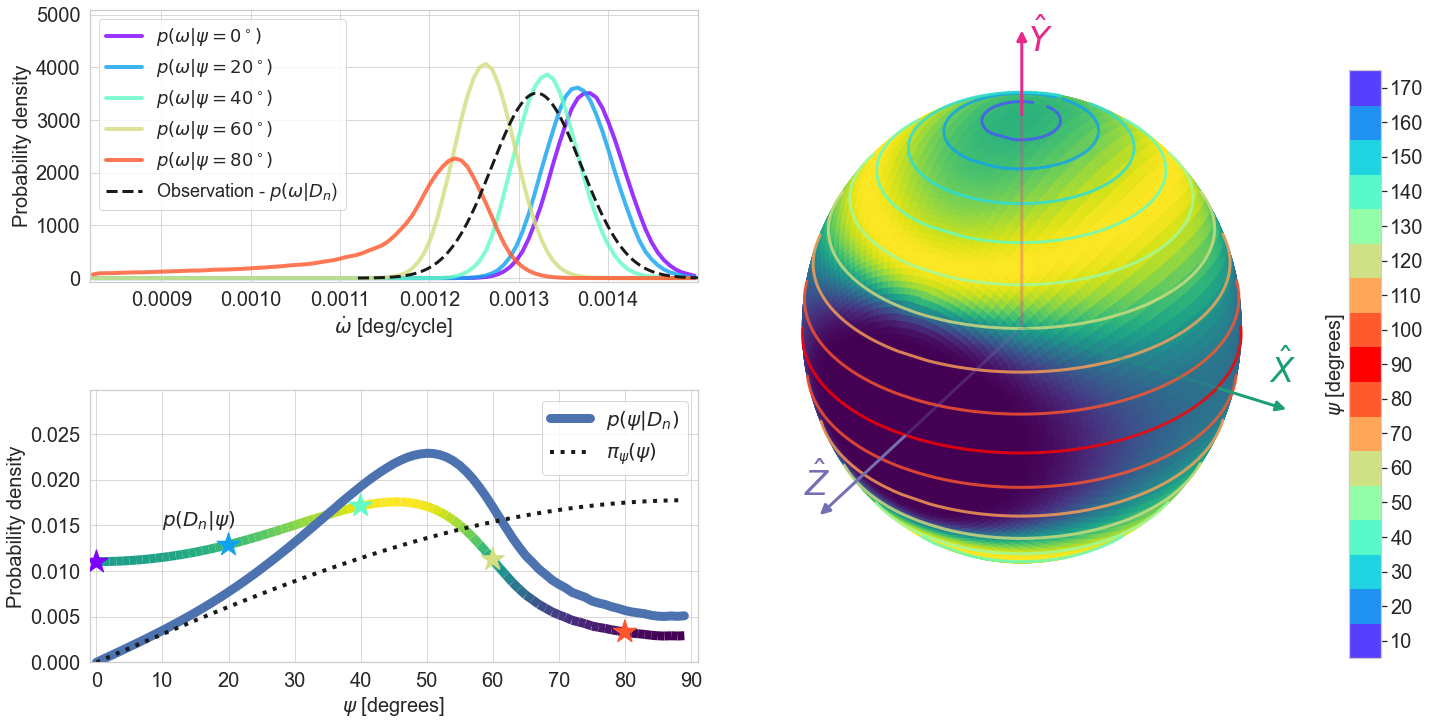} 
\end{center}
\vskip -0.2in
\caption{{\bf VV Pyx.} Using the system VV Pyx as an example, in the top left panel is shown the observed apsidal motion rate $p(\dot \omega|D_n)$ as a black dashed line as well as the distribution $p(\dot \omega|\psi)$ for 5 values of $\psi$. These terms are used in equation \ref{eq:psi likelihood} to calculate $p(D_n|\psi)$, shown in the bottom left panel. On the line of $p(\dot \omega|\psi)$ are shown 5 star symbols that represent the integral of the observation with $p(\dot \omega|\psi)$ (equation \ref{eq:psi likelihood}) using the same 5 example values of $\psi$. Note that the distribution $p(\dot \omega|\psi)$ has a long leftward tail for large $\psi$. This is caused by random samples where $i_s$ is very small and hence the term $\dot \omega_{rot}$ contributes strongly towards retrograde motion. The value $p(D_n|\psi)$ can be thought of as the mean density of $\psi$ when viewed on a sphere as a function of $i_s$ and $\lambda$ corresponding to the representation shown in the right panel where yellow(purple) is higher(lower) probability. Using the uninformative prior for $\pi_\psi (\psi) = \sin \psi$ (dotted black line) gives the probability density $p(\psi|D_n)$ (blue line) as per equation \ref{eq:psi probability}. \label{fig:psi}}
\end{figure*}

$\kappa$ is the concentration parameter of the distribution and when $\kappa \gg 1$, the distribution approaches a Rayleigh distribution. $\kappa$ may then be converted to the scale-parameter of that distribution via the conversion $\sigma = \kappa^{-1/2}$. Given all relevant observations of an ensemble of systems, expressed as the data $D$, the probability distribution of $\kappa$ is given by

\begin{equation} \label{eq:kappaD}
    p(\kappa | D ) = \mathcal{L}_\kappa \pi_\kappa (\kappa) 
\end{equation}

where, for the best comparison with previous studies, we use the same prior $\pi_\kappa (\kappa) = (1 + \kappa^2)^{(-3/4)}$ as \cite{fabrycky_winn2009} and \cite{munoz2018}. The likelihood function of the ensemble is the product of each individual system's likelihood function, given its observed apsidal motion rate:

\begin{equation}
    \mathcal{L}_\kappa = \prod_{n=1}^N \mathcal{L}_{\kappa,n},
\end{equation}

where the likelihood function for a single system is given by

\begin{equation}
    \mathcal{L}_{\kappa,n} \propto \int_0^{\pi} p(\psi | D_n) p(\psi | \kappa)  d\psi.
\end{equation}

here $p(\psi | \kappa)$ is the Fisher distribution and $p(\psi|D_n)$ is the probability distribution of $\psi$ given the observed data of system $n$. $D_n$ comprises the entire data set used to obtain values of the apsidal motion rate as well as stellar radii, masses, internal structure constants, projected rotation speeds and the orbital period of system $n$. As per Bayes's theorem, $p(\psi|D_n)$ is given by: 

\begin{equation} \label{eq:psi probability}
    p (\psi | D_n) \propto p(D_n|\psi) \pi_\psi(\psi)
\end{equation}

where we use the uninformative prior in $\psi$, $\pi_\psi(\psi) = \sin \psi$ (see fig. \ref{fig:psi}) and where $p (D_n | \psi)$ is given by

\begin{equation} \label{eq:psi likelihood}
    p (D_n | \psi) \propto \int_{-\infty}^\infty p(\dot \omega|\psi)p(\dot \omega|D_n)d\dot \omega.
\end{equation}

The term $p(\dot \omega|D_n)$ represents the observed apsidal motion rate $\dot \omega_{obs}$ of a system shown as a black dashed line in the top left panel of fig.~\ref{fig:psi} using the system VV Pyx as an example. For all systems we assume this distribution to be Gaussian with mean and dispersion equal to the quoted values we obtained from the literature. $p(\dot \omega|\psi)$, the PDF of the theoretical apsidal motion rate for a given misalignment $\psi$, is obtained by performing Monte Carlo simulations of equation \ref{eq:omegaTotal} where $\psi$ is fixed at a given value, while drawing from normal distributions of $i_o$, $P$, $e$, $M_j$, $R_j$, $e$, $\log k_{2,j}$ and $v \sin i_{s,j}$. In the top left panel of fig.~\ref{fig:psi}, along with the observed rate, examples of $p(\dot \omega|\psi)$ obtained this way are shown colored by the given obliquity. The bottom panel, displays the three terms from equation \ref{eq:psi probability}.

Given our choice of an uninformative prior, when sampling a given value of $\psi$, $i_{s,j}$, and $\lambda$ must be sampled such that any direction of the stellar rotation axis is as likely as any other i.e.~along a contour line of $\psi$ (denoted the "$\mathbf{n_s}$ ellipse" in \citealt{fabrycky_winn2009}), when we have no information about neither $i_s$ nor $\lambda$. We do however put an upper limit on the rotation velocity of all stars based on the simple expression for the break-up velocity, $v_c = \left( 2GM/3R \right)^{1/2}$. It is worth noting that when $i_o = 90^\circ$, a given value $\dot \omega$ can better constrain $\psi$ than is the case when $i_o \neq 90^\circ$ because a contour of equal $\dot \omega$ crosses more contours of $\psi$. This can be seen when comparing the middle panels to the right panels of fig.~\ref{fig:dependencies}.

Viewing the 48 systems as an ensemble drawn from the same Fisher distribution, together these systems constrain the concentration to $\kappa = 6.7^{+1.9}_{-1.4}$. For a comparison of this concentration with obliquity distribution studies in exoplanetary systems \cite{munoz2018} find $\kappa = 14.5$ for the California-Kepler survey planets. 

\begin{figure}
  \begin{center}
    \includegraphics[trim= 2cm 0.5cm 2cm 2cm, clip,width=1\columnwidth]{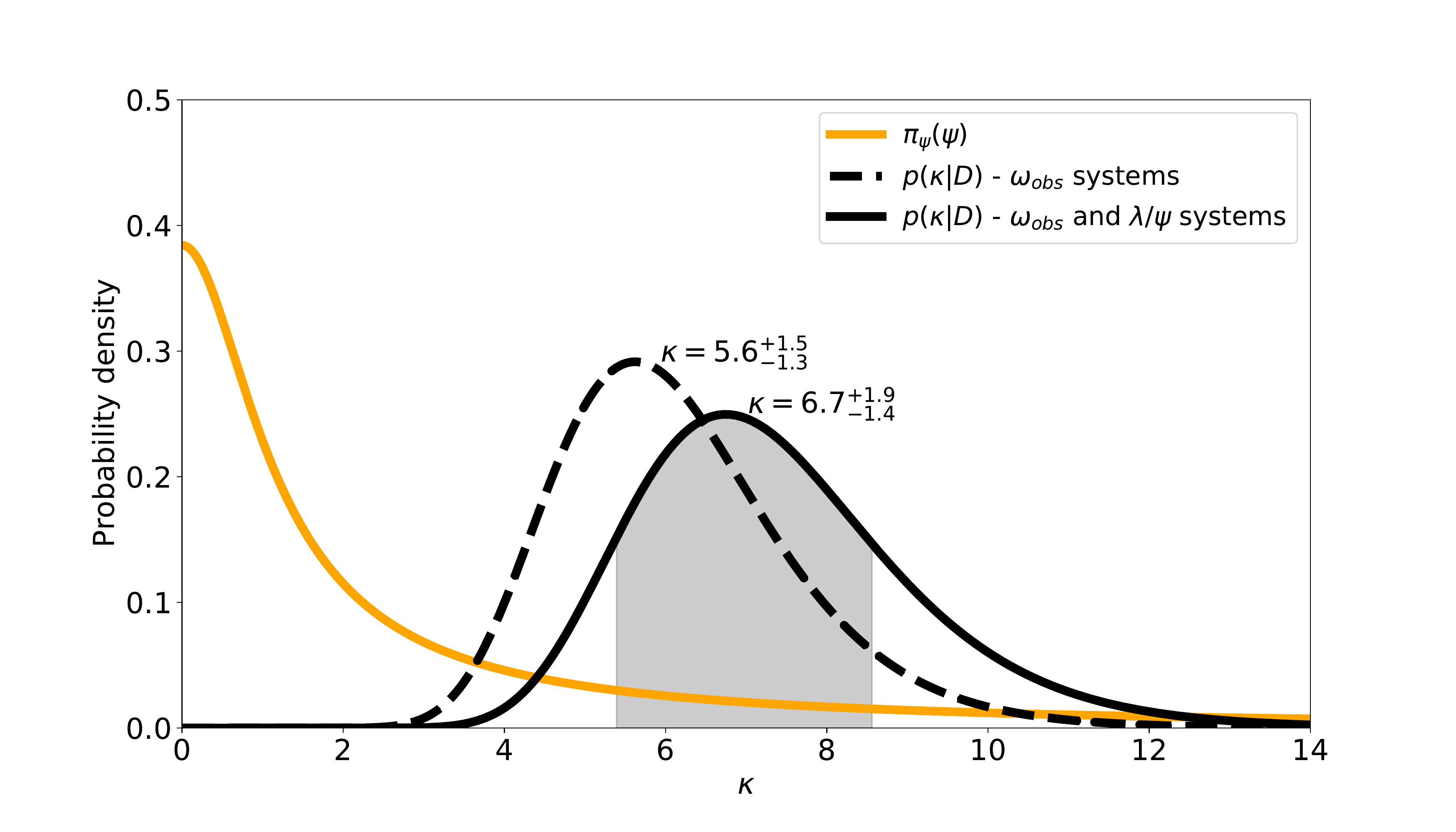}
    \caption{
    \label{fig:kappa posterior} 
    \textbf{Concentration posterior of the ensemble.}
    Shown here are posteriors of the concentration of the two following ensembles, obtained via equation \ref{eq:kappaD}: I) As a dashed line, the 38 aligned systems of table \ref{tab:systems} whose alignment is constrained solely by $\dot \omega_{obs}$. II) As a solid line, the entire sample of table \ref{tab:systems} except DI Her, AS Cam, and CV Vel.
    }
  \end{center}
\end{figure}

\subsection{A subset of perfectly aligned systems?}
The probability distribution of a polar angle goes to zero when the angle goes to zero, as is evident in the right-hand panel of fig.~\ref{fig:All posterios}. Looking at these probability distributions alone, many may seem inconsistent with probability distributions expected for aligned systems. However, this may simply be a consequence of our poor ability to constrain the alignment, resulting in a large dispersion of the distribution on the sphere in the observer-oriented coordinate system. To test the hypothesis that all the systems excluding CV Vel, DI Her, and AS Cam are drawn from an underlying distribution of perfectly aligned systems, we employ a simple bootstrap resampling technique, where we assume the systems to be perfectly spin-orbit aligned, $\psi_j =0$~deg. Assuming Gaussian uncertainties we calculate the relevant physical parameters, $i_o$, $P$, $R$, $M$, $e$, $v \sin i_s$, $\log{k_2}$, using their quoted means and uncertainties. From these values we can then calculate an expected apsidal motion rate, equation \ref{eq:omegaTotal}.

We also need to recalculate the observed apsidal motion rate for each system so it corresponds to the rate of a perfectly aligned system. For this we use equation \ref{eq:omegaTotal} using $\psi_1 = \psi_2 = 0^\circ$ and the system parameters for that system. We then assign the uncertainty from the actual $\dot \omega$ measurement to this new observed rate.

For each draw we then carry out the same comparison between the expected and measured apsidal motion rates as done before in section~\ref{sec:ensemble}. Also as in that section, if no prior knowledge is available for $\lambda$ and $\cos i_s$, then these are obtained from uniform distributions.

We find that  $66\%$ of all draws from our perfectly aligned ensembles have a smaller $\kappa$  - larger dispersion in $\psi$ - than our result of $\kappa = 6.7$ on the actual sample. In this sense our data is fully consistent with an underlying distribution of close binary systems whose stellar and orbital spins are perfectly aligned. 
The somewhat low concentration we measure can thus be explained by measurement uncertainties alone and $\kappa = 6.7$ may be considered reflecting the minimum but not necessarily the actual degree of alignment in this sample.

\section{Summary} \label{sec:discussion}

Combining 35 new, precise apsidal motion rates from \cite{baroch2021} and \cite{Claret+2021} with 16 additional data points of apsidal motion rates as well as measurements of the RM effect and gravity darkening from the literature, we are able to find consistent results of misalignment for the two binary systems DI Her and CV Vel where misalignment is already established. For AS Cam, we agree on the hypothesis by \cite{pavlovski2011} that the low apsidal motion rate of AS Cam can be explained by spin-orbit misalignment alone. 

Among the remaining 48 binaries no misalignment has previously been reported on, and indeed we find no system inconsistent with alignment based on their apsidal motion rates. We found that the dispersion of these systems around $\psi = 0^\circ$  can be described by a Von-Mises Fisher distribution with a concentration parameter of $\kappa = 6.7^{+1.9}_{-1.4}$. A bootstrap experiment then suggests that this dispersion might be attributed to measurement uncertainties alone. Therefore in order to distinguish the distribution of this ensemble from that of 48 perfectly aligned systems, more precise stellar parameters are needed when applying the method described here.

Given the still small number of misaligned systems and wide range of stellar and orbital parameters covered by the 51 systems, we defer any discussion of potential trends until more obliquity measurements for close double star systems become available. As part of the BANANA survey we are measuring obliquities in additional systems.

\newpage

\startlongtable
\begin{deluxetable*}{l LL LLL LL  LL LL}% DD DD}
\tablecaption{\label{tab:systems} Listing of the system and stellar parameters of double star systems in this study. The parameters of the primary component are listed on the same line as the system name and the parameters of the secondary component are listed on the line below the system name. We rounded here $P_o,i_o,e,M,R$, and $\log K_2$ and omitted uncertainties. These are provided in a machine-readable table available for download. For the systems Kepler-16 and KOI-368 we inferred inclinations using the approach of \cite{MasudaWinn2020} and with these inclinations determined obliquities using the method described in \cite{AlbrechtMarcussenWinn+2021}. System parameters have been obtained from the references listed in the notes. We note that the mass of the secondary in the system KOI-368 is not well known and may be around the brown dwarf limit.}
\tablewidth{0pt}
\tablehead{
\colhead{System} 
& $P_o$  & $i_o$ & $e$ & $M$ & $R$ &  \log{k_{2}} & $\dot{\omega}_{\rm obs}$  & $v \sin i_{s}$& $\lambda$ & $i_{s}$ & $\langle \psi \rangle / \psi_j$\\
\colhead{} 
& (\mathrm{days})  &($^\circ$) &  & (\mathrm{M}$_\odot$) & (\mathrm{R}$_\odot$) &   & ($''/cycle$) & \mathrm{(km\,s}$^{-1}$) & ($^\circ$)& ($^\circ$)& ($^\circ$)}

\startdata
 EP Cru$^{1,4}$    & 11.1  & 89.7 & 0.19 & 5.02  & 3.59 & -2.35  & 8.53 \pm 0.54   & 141 \pm 5     & -1.8 \pm 1.6  &                & $6.0^{+7.1}_{-4.5}$    \\
                   &       &      &      & 4.83  & 3.50 & -2.331 &                 & 138 \pm 5     & 0.0 \pm 10.0  &                &                        \\
 Kepler-16$^{15}$  & 41.0  & 89.3 & 0.16 & 0.69  & 0.65 &        &                 & 0.9 \pm 0.03  & 1.6 \pm 2.4   & 90 $\pm 9$     & $8.4^{+10.2}_{-5.2}$   \\
                   &       &      &      & 0.20  & 0.23 &        &                 & 0.9 \pm 0.03  &               &                &                        \\
 KOI-368$^{14}$ & 110.3 & 87.6 & 0.00 & 1.25  & 2.28 &        &                 & 90 \pm 5      & 10.0 \pm 2.0  & 87 $\pm 7$     & $12.9^{+6.2}_{-4.8}$   \\
                   &       &      &      & 0.31  &      &        &                 &               &               &                &                        \\
 AI Hya$^{4}$      & 8.3   & 89.3 & 0.23 & 2.14  & 3.96 & -2.68  & 6.88 \pm 0.18   & 40 \pm 10     &               &                & $14.1^{+11.1}_{-7.4}$  \\
                   &       &      &      & 1.97  & 2.81 &        &                 & 29 \pm 10     &               &                &                        \\
 V1647 Sgr$^{4}$   & 3.3   & 90.0 & 0.41 & 2.18  & 1.83 & -2.38  & 19.94 \pm 0.18  & 80 \pm 5      &               &                & $19.3^{+7.8}_{-11.7}$  \\
                   &       &      &      & 1.97  & 1.67 &        &                 & 70 \pm 5      &               &                &                        \\
 U Oph$^{8}$       & 1.7   & 87.9 & 0.00 & 5.09  & 3.48 & -2.28  & 285.08 \pm 0.04 & 110 \pm 6     &               &                & $16.9^{+13.7}_{-8.7}$  \\
                   &       &      &      & 4.58  & 3.08 &        &                 & 108 \pm 6     &               &                &                        \\
 Y Cyg$^{4}$       & 3.0   & 86.5 & 0.15 & 17.73 & 5.82 & -1.94  & 222.41 \pm 0.11 & 147 \pm 10    &               &                & $21.7^{+11.2}_{-12.2}$ \\
                   &       &      &      & 17.72 & 5.79 &        &                 & 138 \pm 10    &               &                &                        \\
 QX Car$^{4}$      & 4.5   & 85.7 & 0.28 & 9.25  & 4.29 & -2.11  & 43.99 \pm 0.79  & 120 \pm 10    &               &                & $22.6^{+12.2}_{-12.0}$ \\
                   &       &      &      & 8.46  & 4.05 &        &                 & 110 \pm 10    &               &                &                        \\
 CW Cep$^{4}$      & 2.7   & 81.8 & 0.03 & 12.95 & 5.52 & -2.06  & 209.88 \pm 1.44 & 105.2 \pm 2.1 &               &                & $23.2^{+12.9}_{-12.5}$ \\
                   &       &      &      & 11.88 & 5.09 &        &                 & 96.2 \pm 1.9  &               &                &                        \\
 V1143 Cyg$^{2,4}$ & 7.6   & 89.3 & 0.54 & 1.36  & 1.35 & -2.24  & 2.88 \pm 0.14   & 19.6 \pm 0.1  & 7.0 \pm 6.0   &                & $15.9^{+21.7}_{-8.5}$  \\
                   &       &      &      & 1.33  & 1.32 &        &                 & 28.2 \pm 0.1  & -2.0 \pm 3.0  &                &                        \\
 IM Per$^{4}$      & 2.3   & 84.4 & 0.05 & 1.78  & 2.41 & -2.62  & 52.56 \pm 1.44  & 57.5 \pm 3    &               &                & $30.4^{+14.7}_{-15.3}$ \\
                   &       &      &      & 1.77  & 2.37 &        &                 & 56.5 \pm 3    &               &                &                        \\
 NY Cep$^{10}$     & 15.3  & 78.8 & 0.44 & 10.70 & 6.00 &        &                 & 78 \pm 3      & 2.0 \pm 4.0   &                & $9.9^{+33.8}_{-6.3}$   \\
                   &       &      &      & 8.80  & 5.80 &        &                 & 155 \pm 6     &               &                &                        \\
 V541 Cyg$^{5}$    & 15.3  & 89.8 & 0.47 & 2.33  & 1.86 & -2.37  & 1.27 \pm 0.01   & 15 \pm 1      &               &                & $31.3^{+15.4}_{-15.9}$ \\
                   &       &      &      & 2.26  & 1.81 & -2.34  &                 & 15 \pm 1      &               &                &                        \\
 FM Leo$^{11}$     & 6.7   & 88.0 & 0.00 & 1.32  & 1.65 &        &                 & 22 \pm 9      & 0.0 \pm 1.1   &                & $8.1^{+36.2}_{-7.2}$   \\
                   &       &      &      & 1.29  & 1.51 &        &                 & 13 \pm 7.3    & -0.3 \pm 27.4 &                &                        \\
 NN Del$^{11}$     & 99.3  & 89.6 & 0.52 & 1.33  &      &        &                 & 18 \pm 8      & 0.0 \pm 2.6   &                & $8.1^{+37.5}_{-5.6}$   \\
                   &       &      &      & 1.45  &      &        &                 &               &               &                &                        \\
 V963 Cen$^{11}$   & 15.3  &      & 0.42 &       &      &        &                 & 15 \pm 11     & 0.0 \pm 2.4   &                & $8.1^{+37.5}_{-5.7}$   \\
                   &       &      &      &       &      &        &                 &               &               &                &                        \\
 GG Lup$^{4}$      & 1.8   & 86.7 & 0.15 & 4.11  & 2.38 & -2.20  & 62.28 \pm 1.08  & 97 \pm 8      &               &                & $30.4^{+17.6}_{-16.9}$ \\
                   &       &      &      & 2.50  & 1.76 &        &                 & 61 \pm 5      &               &                &                        \\
 V364 Lac$^{4}$    & 7.4   & 89.3 & 0.29 & 2.33  & 3.31 & -2.61  & 6.52 \pm 0.22   & 45 \pm 1      &               &                & $30.4^{+19.4}_{-17.0}$ \\
                   &       &      &      & 2.29  & 2.99 &        &                 & 15 \pm 1      &               &                &                        \\
 Zeta Phe$^{4}$    & 1.7   & 89.1 & 0.01 & 3.91  & 2.83 & -2.29  & 118.08 \pm 2.16 & 85 \pm 8      &               &                & $30.7^{+19.4}_{-17.0}$ \\
                   &       &      &      & 2.54  & 1.89 &        &                 & 75 \pm 8      &               &                &                        \\
 TOI-1338$^{12}$   & 14.6  & 89.7 & 0.16 & 1.13  & 1.33 &        &                 & 3.6 \pm 0.6   & 2.9 \pm 16.0  &                & $25.6^{+26.3}_{-15.0}$ \\
                   &       &      &      & 0.31  & 0.31 &        &                 &               &               &                &                        \\
 V530 Ori$^{5}$    & 6.1   & 89.8 & 0.09 & 1.00  & 0.98 & -0.80  & 3.1 \pm 0.18    & 9 \pm 1       &               &                & $37.0^{+16.1}_{-18.8}$ \\
                   &       &      &      & 0.60  & 0.59 &        &                 & 5 \pm 1       &               &                &                        \\
 V459 Cas$^{5}$    & 8.5   & 89.5 & 0.02 & 2.02  & 2.01 & -2.48  & 2.34 \pm 0.36   & 54 \pm 2      &               &                & $32.5^{+19.8}_{-18.0}$ \\
                   &       &      &      & 1.96  & 1.97 & -2.47  &                 & 43 \pm 2      &               &                &                        \\
 V578 Mon$^{4}$    & 2.4   & 72.1 & 0.08 & 14.54 & 5.41 & -2.00  & 255.2 \pm 0.76  & 117 \pm 4     &               &                & $38.2^{+15.6}_{-19.6}$ \\
                   &       &      &      & 10.29 & 4.29 &        &                 & 94 \pm 2      &               &                &                        \\
 PV Cas$^{4}$      & 1.8   & 85.8 & 0.03 & 2.82  & 2.30 & -2.38  & 76.32 \pm 0.72  & 67 \pm 5      &               &                & $39.7^{+16.7}_{-20.2}$ \\
                   &       &      &      & 2.76  & 2.26 &        &                 & 66 \pm 5      &               &                &                        \\
 V478 Cyg$^{4}$    & 2.9   & 78.2 & 0.02 & 15.40 & 7.26 & -2.24  & 376.92 \pm 3.6  & 129.1 \pm 3.6 &               &                & $41.5^{+17.5}_{-22.9}$ \\
                   &       &      &      & 15.02 & 7.15 &        &                 & 127 \pm 3.5   &               &                &                        \\
 PT Vel$^{4}$      & 1.8   & 88.2 & 0.11 & 2.20  & 2.09 & -2.45  & 45 \pm 2.16     & 63 \pm 2      &               &                & $40.6^{+22.0}_{-22.5}$ \\
                   &       &      &      & 1.63  & 1.56 &        &                 & 40 \pm 3      &               &                &                        \\
 V636 Cen$^{7}$    & 4.3   & 89.6 & 0.13 & 1.05  & 1.02 & -1.61  & 2.88 \pm 0.18   & 13 \pm 0.2    &               &                & $46.4^{+20.0}_{-24.0}$ \\
                   &       &      &      & 0.85  & 0.83 &        &                 & 11.2 \pm 0.5  &               &                &                        \\
 IQ Per$^{4}$      & 1.7   & 89.3 & 0.07 & 3.50  & 2.44 & -2.30  & 54 \pm 1.8      & 68 \pm 2      &               &                & $44.2^{+23.0}_{-23.4}$ \\
                   &       &      &      & 1.73  & 1.50 &        &                 & 44 \pm 2      &               &                &                        \\
 IT Cas$^{5}$      & 3.9   & 89.6 & 0.09 & 1.33  & 1.59 & -2.45  & 4.1 \pm 0.36    & 19 \pm 2      &               &                & $43.3^{+24.4}_{-22.4}$ \\
                   &       &      &      & 1.33  & 1.33 &        &                 & 17 \pm 2      &               &                &                        \\
 EK Cep$^{4}$      & 4.4   & 90.0 & 0.11 & 2.02  & 1.58 & -2.17  & 3.17 \pm 0.14   & 23 \pm 2      &               &                & $46.7^{+22.4}_{-23.9}$ \\
                   &       &      &      & 1.12  & 1.31 &        &                 & 10.5 \pm 2    &               &                &                        \\
 V760 Sco$^{4}$    & 1.7   & 82.1 & 0.03 & 4.97  & 3.02 & -2.18  & 156.24 \pm 1.8  & 95 \pm 10     &               &                &                        \\
                   &       &      &      & 4.61  & 2.64 &        &                 & 85 \pm 10     &               &                &                        \\
 VV Pyx$^{4}$      & 4.6   & 88.0 & 0.10 & 2.10  & 2.17 & -2.49  & 4.75 \pm 0.18   & 23 \pm 3      &               &                &                        \\
                   &       &      &      & 2.10  & 2.17 &        &                 & 23 \pm 3      &               &                &                        \\
 V539 Ara$^{4}$    & 3.2   & 85.1 & 0.05 & 6.24  & 4.52 & -2.31  & 70.56 \pm 2.16  & 75 \pm 8      &               &                &                        \\
                   &       &      &      & 5.31  & 3.43 &        &                 & 48 \pm 5      &               &                &                        \\
 AG Per$^{7}$      & 2.0   & 81.4 & 0.07 & 5.35  & 3.00 & -2.14  & 94.32 \pm 1.44  & 94 \pm 23     &               &                &                        \\
                   &       &      &      & 4.89  & 2.60 &        &                 & 70 \pm 9      &               &                &                        \\
 SW CMa$^{4}$      & 10.1  & 88.6 & 0.32 & 2.24  & 3.01 & -2.58  & 2.48 \pm 0.18   & 24 \pm 1.5    &               &                &                        \\
                   &       &      &      & 2.10  & 2.40 &        &                 & 10 \pm 1      &               &                &                        \\
 EY Cep$^{5}$      & 8.0   & 89.9 & 0.44 & 1.52  & 1.46 & -2.38  & 1.83 \pm 0.06   & 10 \pm 1      &               &                &                        \\
                   &       &      &      & 1.50  & 1.47 & -2.4   &                 & 10 \pm 1      &               &                &                        \\
 EM Car$^{4}$      & 3.4   & 81.5 & 0.01 & 22.83 & 9.35 & -2.26  & 288 \pm 7.2     & 150 \pm 20    &               &                &                        \\
                   &       &      &      & 21.38 & 8.35 &        &                 & 130 \pm 15    &               &                &                        \\
 V453 Cyg$^{4}$    & 3.9   & 87.5 & 0.03 & 13.96 & 8.67 & -2.38  & 155.16 \pm 5.4  & 107.2 \pm 2.8 &               &                &                        \\
                   &       &      &      & 11.10 & 5.25 &        &                 & 98.3 \pm 3.7  &               &                &                        \\
 V335 Ser$^{6}$    & 3.4   & 87.2 & 0.14 & 2.15  & 2.03 & -2.44  & 13.94 \pm 5.44  & 30 \pm 2      &               &                &                        \\
                   &       &      &      & 1.91  & 1.73 &        &                 & 51 \pm 3      &               &                &                        \\
 KX Cnc$^{5}$      & 31.2  & 89.8 & 0.47 & 1.14  & 1.06 & -1.88  & 0.47 \pm 0.04   & 6.4 \pm 1     &               &                &                        \\
                   &       &      &      & 1.13  & 1.05 &        &                 & 6.5 \pm 1     &               &                &                        \\
 GG Ori$^{5}$      & 6.6   & 89.3 & 0.22 & 2.34  & 1.85 & -2.34  & 2.2 \pm 0.11    & 24 \pm 2      &               &                &                        \\
                   &       &      &      & 2.34  & 1.83 & -2.337 &                 & 23 \pm 2      &               &                &                        \\
 VV Crv$^{4}$      & 3.1   & 88.9 & 0.09 & 1.98  & 3.38 & -2.69  & 39.24 \pm 4.32  & 81 \pm 3      &               &                &                        \\
                   &       &      &      & 1.51  & 1.65 &        &                 & 24 \pm 2      &               &                &                        \\
 KW Hya$^{5}$      & 7.8   & 87.5 & 0.09 & 1.96  & 2.12 & -2.50  & 1.62 \pm 0.25   & 15 \pm 2      &               &                &                        \\
                   &       &      &      & 1.49  & 1.44 & -2.45  &                 & 13 \pm 2      &               &                &                        \\
 V451 Oph$^{7}$    & 2.2   & 85.9 & 0.01 & 2.77  & 2.64 & -2.45  & 43.2 \pm 2.52   & 41 \pm 7      &               &                &                        \\
                   &       &      &      & 2.35  & 2.03 &        &                 & 30 \pm 5      &               &                &                        \\
 V501 Mon$^{5}$    & 7.0   & 88.0 & 0.13 & 1.65  & 1.89 & -2.54  & 1.66 \pm 0.4    & 16.5 \pm 1    &               &                &                        \\
                   &       &      &      & 1.46  & 1.59 & -2.52  &                 & 12.4 \pm 1    &               &                &                        \\
 V501 Her$^{5}$    & 8.6   & 89.1 & 0.10 & 1.27  & 2.00 & -2.13  & 1.48 \pm 0.72   & 14.3 \pm 1    &               &                &                        \\
                   &       &      &      & 1.21  & 1.51 & -1.96  &                 & 12.5 \pm 1    &               &                &                        \\
 J1219–39$^{13}$   & 6.8   & 87.6 & 0.00 & 0.83  & 0.81 &        &                 & 2.6 \pm 0.42  & -4.1 \pm 5.3  &                &                        \\
                   &       &      &      & 0.09  & 0.12 &        &                 & 0.9 \pm 0.03  & 4.1 \pm 5.3   &                &                        \\
 AL Dor$^{5}$      & 14.9  & 88.8 & 0.20 & 1.10  & 1.12 & -1.95  & 0.59 \pm 0.02   & 4.6 \pm 1     &               &                &                        \\
                   &       &      &      & 1.10  & 1.12 &        &                 & 4.6 \pm 1     &               &                &                        \\
 CV Vel$^{9}$      & 6.9   & 86.5 & 0.00 & 6.09  & 4.09 & -2.20  &                 & 21.5 \pm 2    & -52.0 \pm 6.0 &                & ^*64 $\pm$ 4           \\
                   &       &      &      & 5.98  & 3.95 & -2.328 &                 & 21.1 \pm 2    & 3.7 \pm 7.0   &                & ^*46 $\pm$ 9           \\
 DI Her$^{1,4}$    & 10.6  & 89.1 & 0.49 & 5.10  & 2.68 & -2.13  & 1.38 \pm 0.02   & 108 \pm 4     & -74.0 \pm 2.0 & $74^{+6}_{-5}$ & $^*74^{+2}_{-3}$       \\
                   &       &      &      & 4.40  & 2.48 &        &                 & 116 \pm 4     & 79.0 \pm 3.0  & 109 $\pm 7$    & $^*80^{+2}_{-3}$       \\
 AS Cam$^{3}$      & 3.4   & 89.5 & 0.16 & 3.21  & 2.60 & -1.55  & 4.5 \pm 0.34    & 14.5 \pm 0.1  &               &                & $87.9^{+2.1}_{-2.5}$   \\
                   &       &      &      & 2.32  & 1.96 & -1.5   &                 & 4.6 \pm 0.1   &               &                &                        \\
\enddata
\tablecomments{If one value is quoted for $\psi$ it is the mode and highest density intervals of the $\langle \psi \rangle$ posteriors. If two values are given, they refer to $\psi_1$ and $\psi_2$. Systems where the data does not constrain $\langle \psi \rangle$ significantly, which we define as a highest density interval longer than $46.4^\circ$, are not included. For some systems $\psi$ for specific components are available in the literature - these $\psi$ values from the literature are marked with a $^*$ symbol. $i_{s,j}$, $\lambda_{j}$, and $i_o$ are functions of time for spin-orbit misaligned systems due to spin and orbital precession. The quoted values for DI Her are valid for June 2020 and CV Vel for 2009-2010. \\
\\
References: 1 \citep{albrecht2013}, 2 \citep{albrecht2007}, 3 \citep{kozyreva2018evolution,pavlovski2011,maloney1989}, 4 \citep{Claret+2021}, 5 \citep{baroch2021}, 6 \citep{SandbergV335}, 7 \citep{torres2010}, 8 \citep{Johnson2019Uoph}, 9 \citep{Albrecht2014CVVel}, 10 \citep{Albrecht2011NYCep}, 11 \citep{sybilski2018}, 12 \citep{Kunovac2020TOI-1338}, 13 \citep{Triaud2013}, 14 \citep{Ahlers2014}, 15 \citep{Winn2011Kepler-16}, 16 (Liang et al.\ submitted).}
\end{deluxetable*}

\begin{acknowledgments}
We thank Tsevi Mazeh for discussions at the onset of this project. We thank Yan Liang for her explanation of equation $\ref{eq:phi}$.
Funding for the Stellar Astrophysics Centre is provided by The Danish National Research Foundation (Grant agreement no.: DNRF106)
\end{acknowledgments}

\bibliography{refs,refs_tools}{}
\bibliographystyle{aasjournal-hyperref}

\end{document}